\documentclass[10pt]{article}
\usepackage{graphicx}

\def\Title#1{\begin{center} {\Large #1 } \end{center}}
\def\Author#1{\begin{center}{ \sc #1} \end{center}}
\def\Address#1{\begin{center}{ \it #1} \end{center}}

\newcommand\pubblock{\rightline{\begin{tabular}{l} Proceedings of the Fifth Annual LHCP\\ \pubnumber\\
         \pubdate  \end{tabular}}}

\newenvironment{Abstract}{\begin{quotation} \begin{center}
             \large ABSTRACT \end{center}\bigskip
      \begin{center}\begin{large}}{\end{large}\end{center} \end{quotation}}

\newenvironment{Presented}{\begin{quotation} \begin{center}
             PRESENTED AT\end{center}\bigskip
      \begin{center}\begin{large}}{\end{large}\end{center} \end{quotation}}

\def\Acknowledgements{\bigskip  \bigskip \begin{center} \begin{large}
             \bf ACKNOWLEDGEMENTS \end{large}\end{center}}

\def\beq{\begin{equation}}
\def\eeq#1{\label{#1}\end{equation}}
\def\eeqn{\end{equation}}

\def\beqa{\begin{eqnarray}}
\def\eeqa#1{\label{#1}\end{eqnarray}}
\def\eeqan{\end{eqnarray}}

\let\bar=\overbar

\def\Dslash{\not{\hbox{\kern-4pt $D$}}}
\def\dslash{\not{\hbox{\kern-2pt $\del$}}}

\def\msb{{\bar{\ssstyle M \kern -1pt S}}}

\usepackage{color}

\newcommand\pubnumber{ CMS CR-2017/197 }

\newcommand\pubdate{\today}

\def\affiliation{
On behalf of the CMS collaboration \\
Institute of High Energy Physics, Chinese Academy of Sciences\\
100049 Beijing, CHINA}

\newcommand{\Hgg}{\ensuremath{H \rightarrow \gamma\gamma}}
\newcommand{\PW}{\ensuremath{{W}}}
\newcommand{\cPZ}{\ensuremath{{Z}}}
\newcommand{\PH}{\ensuremath{{H}}}
\newcommand{\WH}{\ensuremath{\PW\PH}}
\newcommand{\ZH}{\ensuremath{\cPZ\PH}}
\newcommand{\cPqt}{\ensuremath{{t}}}
\newcommand{\ttH}{\ensuremath{\cPqt\bar{\cPqt}\PH}}
\newcommand{\Pe}{\ensuremath{{e}}}

\newcommand{\Pgam}{\ensuremath{{\gamma}}}

\newcommand{\mgg}{\mathrm{m_{\Pgam\Pgam}}}
\newcommand{\mH}{\mathrm{m_{\PH}}}

\newcommand{\sigmaMoMdecorr}{\sigma_{\mathrm{m}}/\mathrm{m}|_{decorr}}

\newcommand{\Pgm}{\ensuremath{{\mu}}}
\newcommand{\Pg}{\ensuremath{{g}}}
\newcommand{\HZZfl}{\ensuremath{\PH\to\cPZ\cPZ\to4\ell}}
\newcommand{\DMeVbfjj}{\ensuremath{{\mathcal D}_{\rm 2jet}}}
\newcommand{\DMeVbfj}{\ensuremath{{\mathcal D}_{\rm 1jet}}}
\newcommand{\DMeWh}{\ensuremath{{\mathcal D}_{\rm \WH}}}
\newcommand{\DMeZh}{\ensuremath{{\mathcal D}_{\rm \ZH}}}

\newcommand{\Pq}{\ensuremath{{q}}}
\newcommand{\Paq}{\ensuremath{\bar{\Pq}}}
\newcommand{\ttbar}{\ensuremath{\cPqt\bar{\cPqt}}}
\newcommand{\KD}{\ensuremath{{\cal D}^{\rm kin}_{\rm bkg}} }
\newcommand{\mllll}{\ensuremath{m_{4\ell}}}
\newcommand{\muV}{\ensuremath{\mu_{\mathrm{VBF},\mathrm{V\PH}}} }
\newcommand{\muF}{\ensuremath{\mu_{\Pg\Pg\PH,\,\ttbar\PH}} }
\newcommand{\MassDprime}{\mathrm{{\cal D}'_{\rm mass}} }
\newcommand{\Pp}{\ensuremath{{p}}}
\newcommand{\Hllll}{\ensuremath{\PH\to4\ell}}
\newcommand{\pt}{\ensuremath{p_{\mathrm{T}}}}

\newcommand{\valMuAtRunIMass}{\ensuremath{1.05^{+0.19}_{-0.17}}}
\newcommand{\valMuVAtRunIMass}{\ensuremath{0.00}^{+1.37}_{-0.00}}
\newcommand{\valMuFAtRunIMass}{\ensuremath{1.20}^{+0.35}_{-0.31}}
\newcommand{\valMassThreeDRefit}{\ensuremath{125.26 \pm 0.20 (\mathrm{stat.}) \pm 0.08 (\mathrm{sys.})}}

\newcommand{\fbinv}{\mathrm{fb}^{-1}}
\newcommand{\usedLumi}{35.9~\fbinv}

\newcommand{\GeV}{\mathrm{{GeV}}}
\newcommand{\TeV}{\mathrm{{TeV}}}

\begin{document}

\large
\begin{titlepage}
\pubblock

\vfill
\Title{  Higgs boson measurements in high mass resolution channels with CMS  }
\vfill

\Author{ JUNQUAN TAO  }
\Address{\affiliation}
\vfill
\begin{Abstract}

The latest measurements of the Higgs boson properties in both
the $\mathrm{H}\rightarrow\gamma\gamma$ decay channel and
the $\mathrm{H}\rightarrow{\rm Z}{\rm Z}\rightarrow4\ell$ ($\ell={\rm e},\mu$)
decay channel using the proton-proton collision data corresponding to an integrated luminosity
of $\usedLumi$ at $\sqrt{s}=13~\TeV$, including the signal strength relative to
the standard model prediction, signal strength modifiers for
different Higgs production modes, coupling modifiers to fermions and
bosons, and effective coupling modifiers to photons and gluons, are presented.
In addition, dedicated measurements of the Higgs boson's mass,
width, total and differential fiducial cross sections have been summarized.
All results are consistent, within their uncertainties, with the expectations for the SM Higgs boson.

\end{Abstract}
\vfill

\begin{Presented}
The Fifth Annual Conference\\
 on Large Hadron Collider Physics \\
Shanghai Jiao Tong University, Shanghai, China\\
May 15-20, 2017
\end{Presented}
\vfill
\end{titlepage}
\def\thefootnote{\fnsymbol{footnote}}
\setcounter{footnote}{0}
%

\normalsize


\section{Introduction}

The standard model of particle physics (SM)~\cite{Glashow:1961tr,Weinberg:1967tq,Salam:1968rm}
has been very successful in explaining high-energy experimental data. During the
Run~1 period of the CERN LHC, a new particle was discovered by both
ATLAS~\cite{Aad:2012tfa} and CMS~\cite{Chatrchyan:2012xdj} experiments
and the collected experimental evidence is consistent with the particle
being a Higgs boson compatible with the quantum of the scalar field postulated by the
Higgs mechanism~\cite{Englert:1964et,Higgs:1964pj,Guralnik:1964eu}.

$\mathrm{H}\rightarrow\gamma\gamma$ and
$\mathrm{H}\rightarrow{\rm Z}{\rm Z}\rightarrow4\ell$ ($\ell={\rm e},\mu$)
were the most two important channels involved in the discovery of the Higgs boson
and first measurements of its properties.
Despite the small branching ratio predicted by the SM ($\approx$ 0.2\%),
the $\mathrm{H}\rightarrow\gamma\gamma$ decay channel provides a clean final
state with an invariant mass peak that can be reconstructed with high
precision. The $\mathrm{H}\rightarrow{\rm Z}{\rm Z}\rightarrow4\ell$ decay channel
($\ell={\rm e},\mu$) has a large signal-to-background ratio, and the precise
reconstruction of the final-state decay products allows the complete determination of the
kinematics of the Higgs boson.

The latest measurements of properties of the Higgs boson SM(125) in both
the $\mathrm{H}\rightarrow\gamma\gamma$ decay channel~\cite{CMS:2017rli,CMS:2017nyv} and
the $\mathrm{H}\rightarrow{\rm Z}{\rm Z}\rightarrow4\ell$ ($\ell={\rm e},\mu$)
decay channel ~\cite{CMS:2017jkd,CMS:2017pij} are presented, including the signal strength relative to
the standard model prediction, signal strength modifiers for
different Higgs production modes, coupling modifiers to fermions and
bosons, and effective coupling modifiers to photons and gluons.
The analyses use the data collected by the CMS experiment in proton-proton collisions
during the 2016 LHC running period. The data sample corresponds to
an integrated luminosity of $35.9~\mathrm{fb}^{-1}$ at $\sqrt{s}=13~\mathrm{TeV}$.
For both decay channels, categories have been introduced targeting subleading production modes
of the Higgs boson such as vector boson fusion (VBF) and associated
production with a vector boson ($\WH$, $\ZH$)
or top quark pair ($\ttH$). In addition, dedicated measurements of the Higgs boson's mass,
width, total and differential cross sections have been summarized.

\section{$\mathrm{H}\rightarrow\gamma\gamma$}
\label{sec:Hgg}

In the $\mathrm{H}\rightarrow\gamma\gamma$  analysis, a search is made for a fully reconstructed peak
in the diphoton invariant mass distribution in the range 110-150$~\mathrm{GeV}$,
on a large irreducible background from QCD production
of two photons. There is also a reducible background where one
or more of the reconstructed photon candidates originate from
misidentification of jet fragments.
To optimize the photon energy resolution, the energy is corrected for the
containment of the electromagnetic showers in the clustered crystals and
the energy losses of converted photons.
%
A Boosted Decision Tree (BDT) is used to separate prompt photons from
photon candidates satisfying the preselection requirements while
resulting from misidentification of jet fragments.
%
The diphoton vertex assignment relies on a boosted
decision tree, whose inputs are observables related to tracks
recoiling against the diphoton system. The average vertex assignment efficiency
is measured to be 80$\%$.
%
To improve the sensitivity of the analysis, events are classified
targeting different production mechanisms and according to their mass
resolution and predicted signal-to-background ratio. A dedicated diphoton multivariate classifier, implemented as a
boosted-decision tree (BDT), is trained to
evaluate the diphoton mass resolution on a per-event basis and is used
as an ingredient in the categorization.
In total 14 event categories were used, in the following order
based on the additional particles present in the event along with
the diphoton pair: ttHLeptonic, ttHHadronic, ZHLeptonic,
WHLeptonic, VHLeptonicLoose, 3 VBF-tagged categories, VHMET, VHHadronic,
and 4 Untagged categories.
To extract signal events parametric models for signal and background have been
built separately for each category. The signal model is extracted from simulation
as a combination of several gaussian function taking into account different
correction and scale factors. The background models are completely data driven where a
nuisance parameter is set to vary over a set of possible functional forms. This
technique is discussed in details in~\cite{Dauncey:2014xga}.

\begin{figure}[htbp]
 \begin{center}
 {\includegraphics[width=0.29\textwidth]{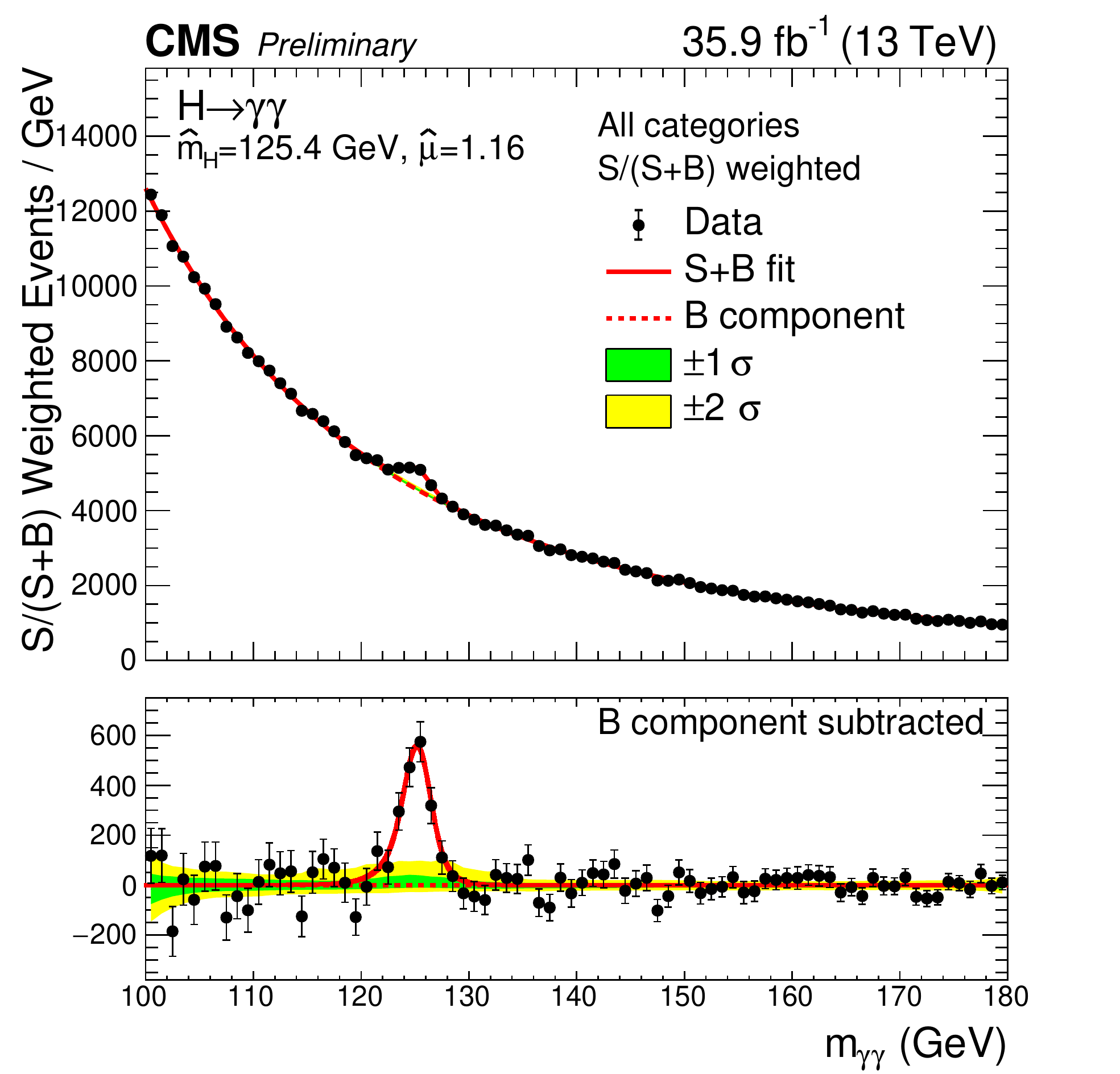}}
  {\includegraphics[width=0.4\textwidth]{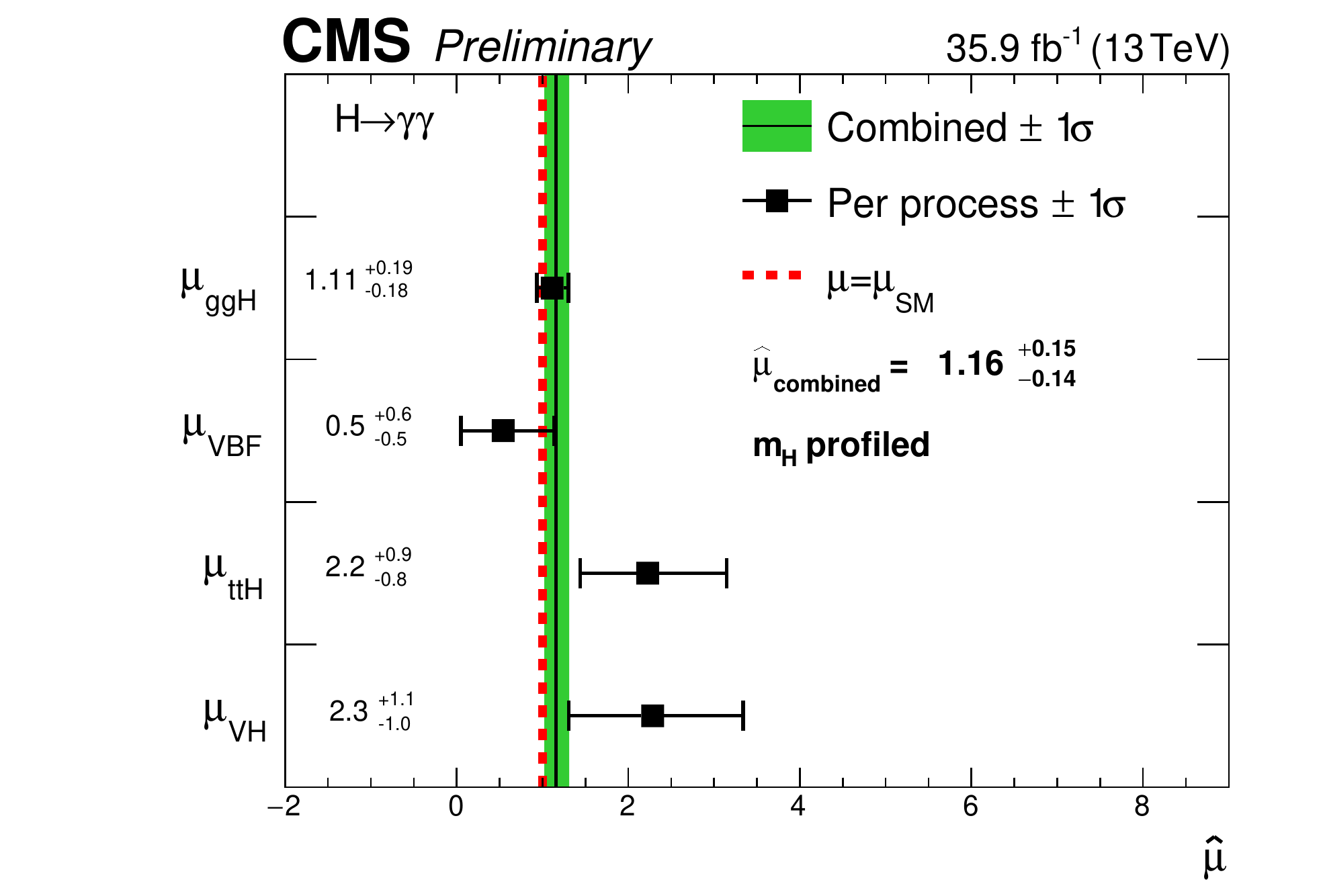}}
    {\includegraphics[width=0.29\textwidth]{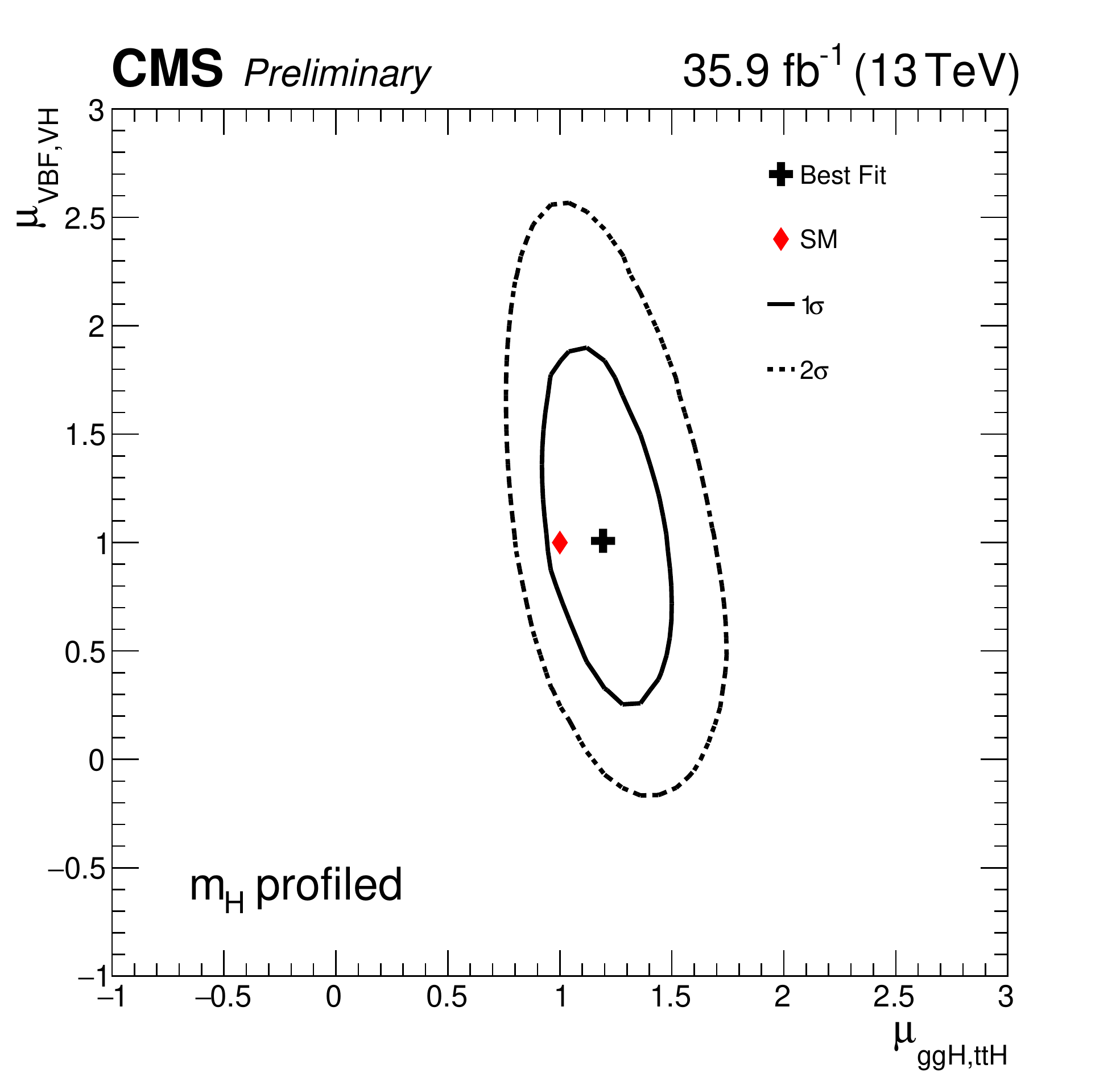}}
  \caption{ (Left) Data points (black) and signal plus background model fits
for all categories summed and weighted by their sensitivity.
The bottom panel shows the residuals after background subtraction.
(Middle) Signal strength modifiers measured for each process (black points)
for profiled $m_{H}$, compared to the overall signal strength
(green band) and to the SM expectation (dashed red line).
(Right) The two-dimensional best-fit (black cross) of the signal
strengths for fermionic (ggH, $\mathrm{t\bar{t}H}$) and bosonic (VBF,
ZH, WH) production modes compared to the SM expectations (red diamond) (right).
The Higgs boson mass is profiled in the fit. }
 \label{fig:Hgg1}
 \end{center}
\end{figure}

Figure~\ref{fig:Hgg1} left shows The $\mgg$ distribution of the sum of all the categories weighted to their
expected signal to signal-plus-background ratio. The one standard deviation (green)
and two standard deviation bands (yellow) include the uncertainties in the
background component of the fit. The best fit value of the signal strength $\mu$ with $\mH$ profiled
is reported to be
$\widehat{\mu} = 1.16^{+0.15}_{-0.14} =
1.16^{+0.11}_{-0.10} ({\rm stat.}) ^{+0.09}_{-0.08} ({\rm sys.}) ^{+0.06}_{-0.05} ({\rm theo.})$ (Figure~\ref{fig:Hgg1} middle).
The best-fit values for the signal strength modifiers associated with the
ggH and $\mathrm{t\bar{t}H}$ production mechanisms, and with the VBF and VH production
processes are measured; the best fit values for each modifier are
$\mu_{ggH,t\bar{t}H} =1.19^{+0.20}_{-0.18}$
and $\mu_{VBF,VH}=1.01^{+0.57}_{-0.51}$ (Figure~\ref{fig:Hgg1} right).

\begin{figure}[h]
\begin{center}
  \includegraphics[width=0.4\textwidth]{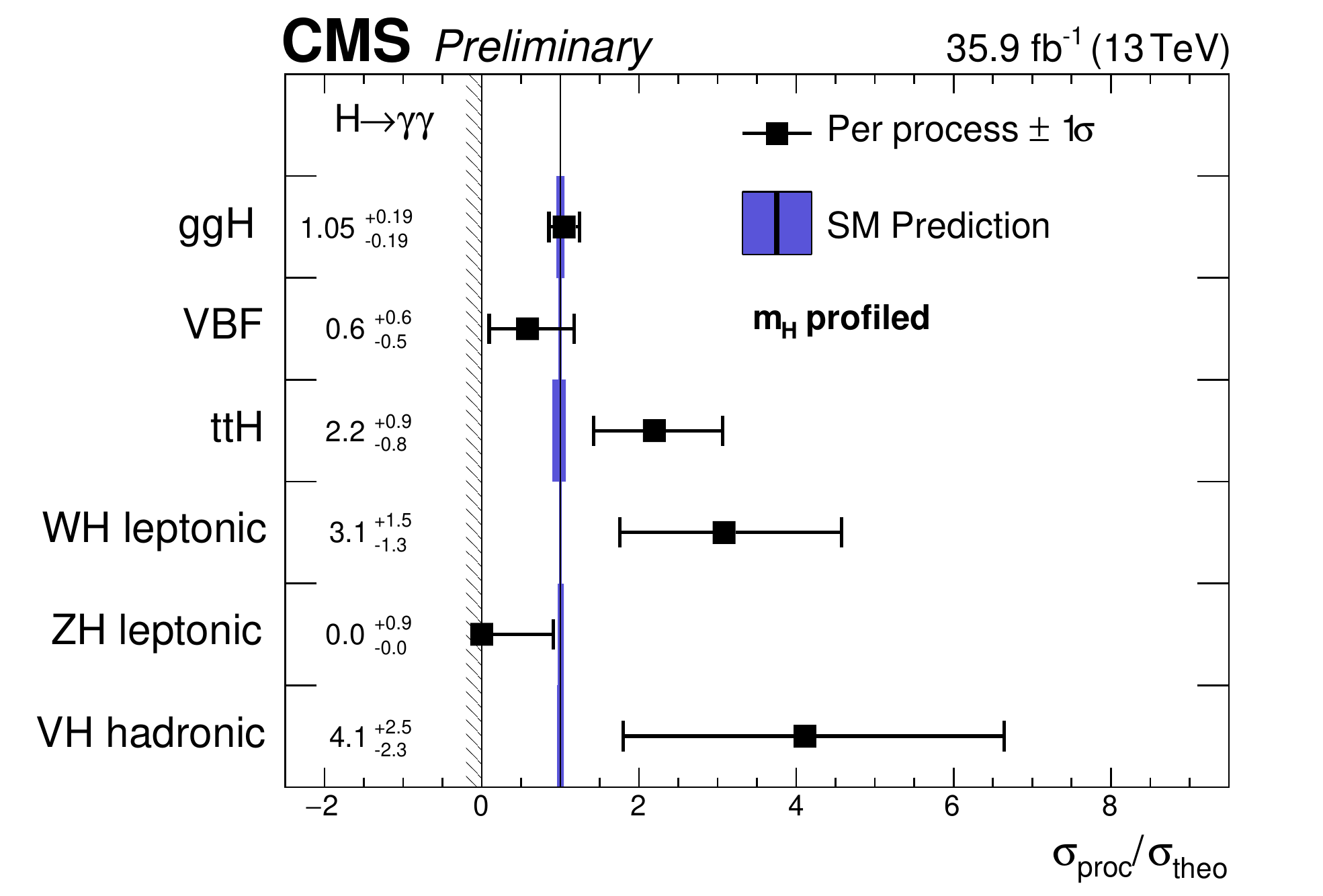}
  {\includegraphics[width=0.29\textwidth]{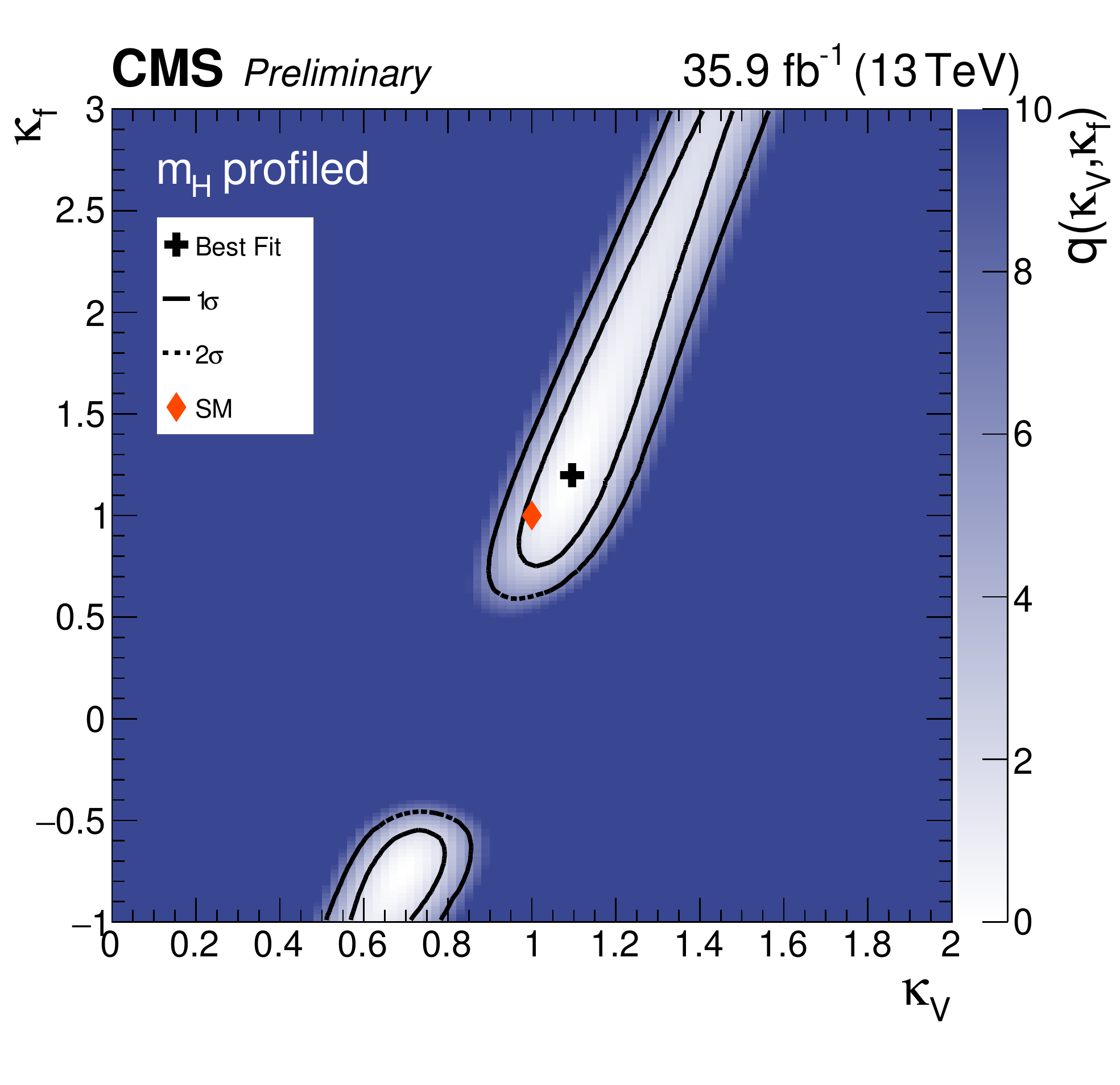}}%
  {\includegraphics[width=0.29\textwidth]{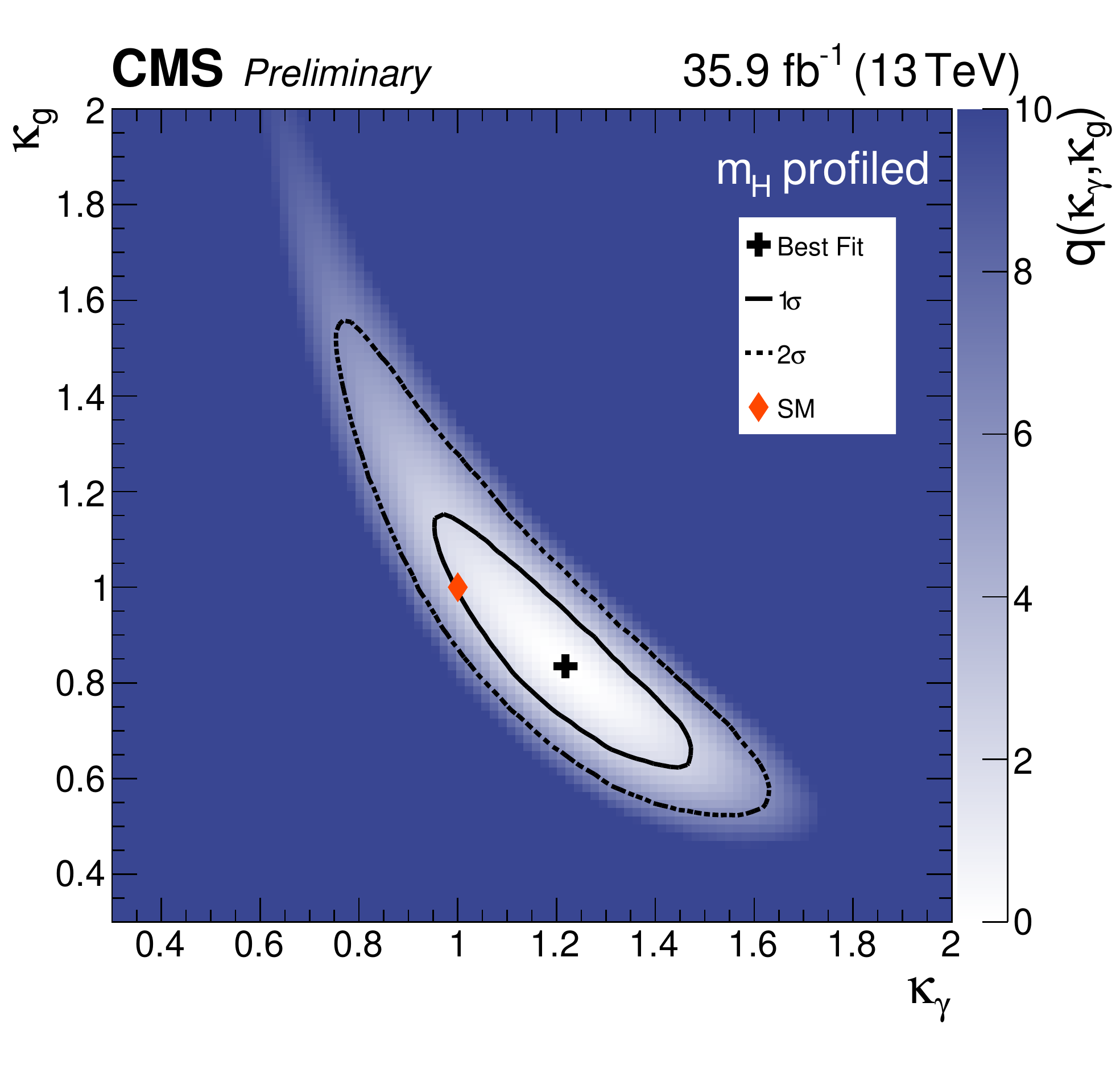}}
\end{center}
\caption{ Cross section ratios measured for each process (black
points) in the Higgs Simplified Template Cross Section framework (left),
for profiled $m_{H}$, compared to the SM expectation and its uncertainties
(green band); Two-dimensional likelihood scans of $\kappa_{f}$ versus $\kappa_{V}$
  (middle) and $\kappa_{g}$ versus $\kappa_{\gamma}$ (right).
}
\label{fig:Hgg2}
\end{figure}

Fig.~\ref{fig:Hgg2} left shows the cross section
ratios measured for each process in the Higgs Simplified Template Cross Section (STXS) framework
detailed in the CERN Yellow Report 4 of the LHC-HXSWG~\cite{deFlorian:2016spz}.
Two-dimensional likelihood scans of the Higgs boson coupling modifiers are produced:
$\kappa_{f}$ versus $\kappa_{V}$, the coupling modifiers to bosons and fermions;
and $\kappa_{\gamma}$ and $\kappa_{g}$, the effective coupling modifiers
to photons and gluons~\cite{Heinemeyer:2013tqa}, as shown in
Fig.~\ref{fig:Hgg2} middle and right with 1$\sigma$ and 2$\sigma$
contours for each scan. All four variables are expressed relative to the SM expectations.
The mass of the Higgs boson is profiled in the fits. The crosses indicate
the best-fit values, the diamonds indicate the Standard Model expectations.

\begin{figure}
 \begin{center}
   \includegraphics[width=0.32\textwidth]{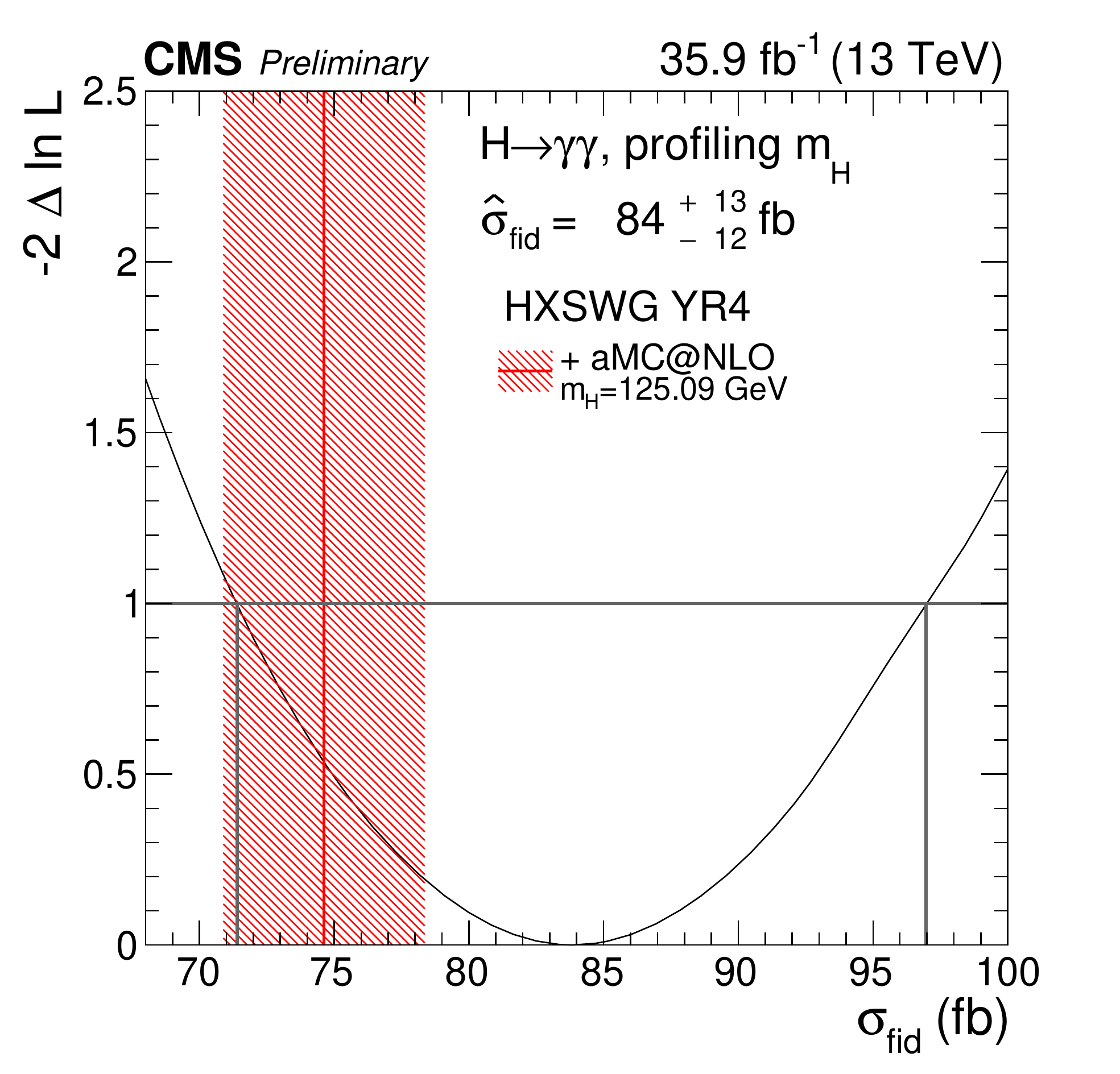}
   \includegraphics[width=0.33\textwidth]{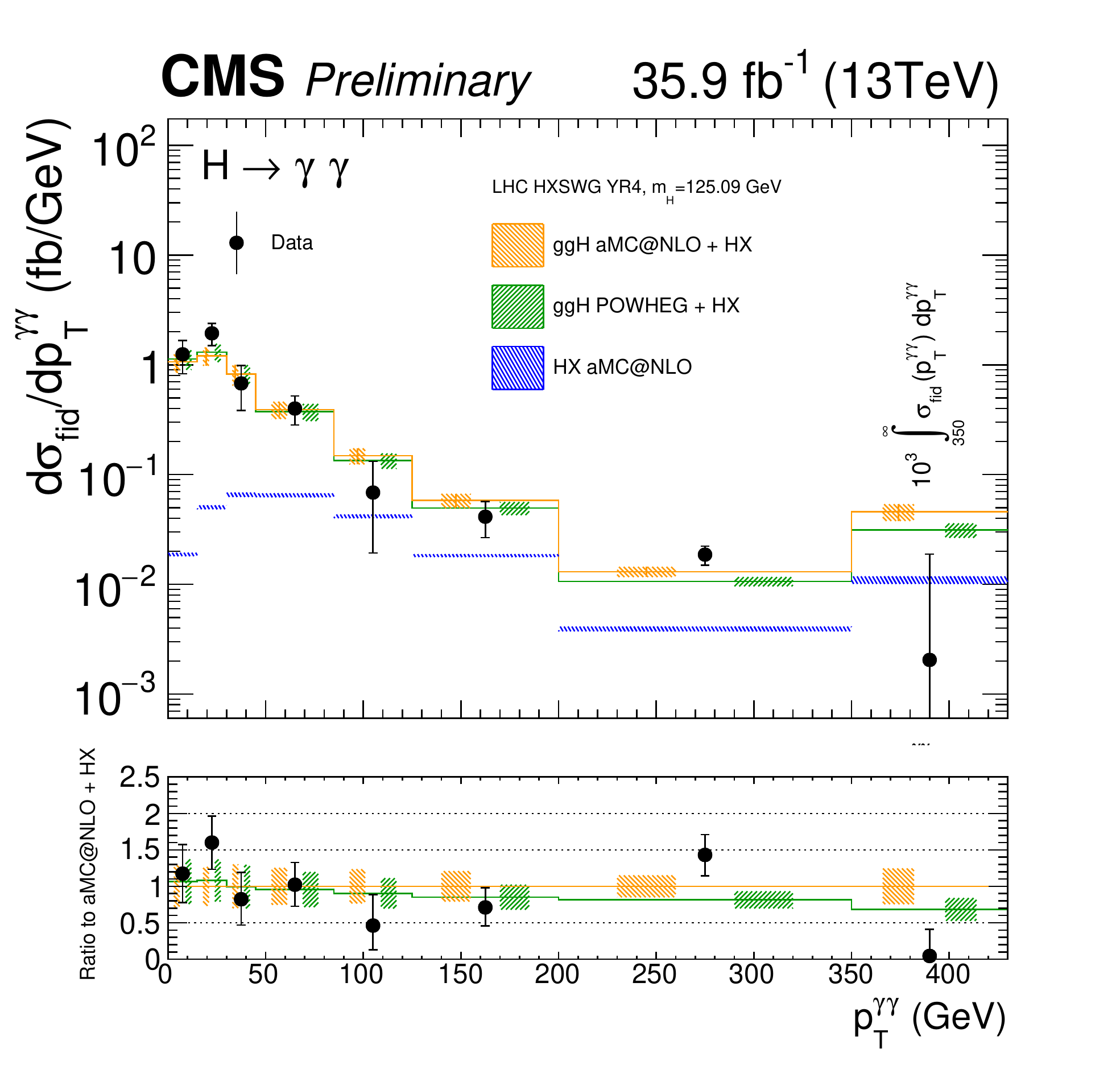}
     \includegraphics[width=0.33\textwidth]{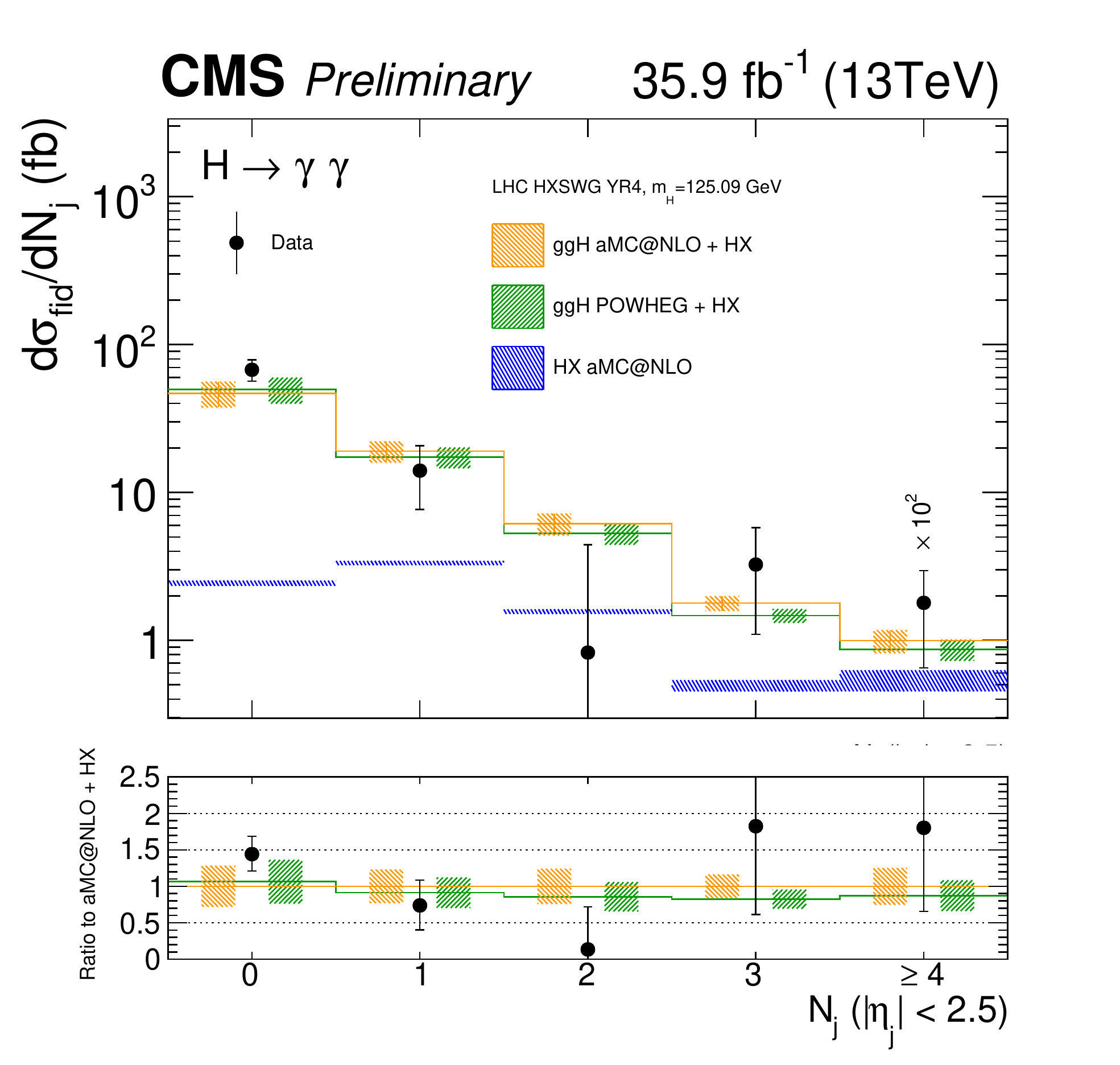}
   \caption{Left plot shows the likelihood scan (black curve) for the fiducial cross section where the value of the SM Higgs boson mass is profiled in the fit. The measurement is compared to the theoretical prediction (in red) and shows good agreement within uncertainties. Measured $\Hgg$ differential cross-section (black points) for $p_{T,\gamma\gamma}$ (middle) and $N_{jets}$ (right) are compared to the theoretical predictions with two different generators for the gluon-gluon fusion process: MADGRAPH\_aMC@NLO (in orange) and POWHEG (in green). The sum of the contributions from VBF, VH and ttH processes, labeled as HX, is generated using MADGRAPH$\_$aMC@NLO and is shown in blue in the plot.}
  \label{fig:Hggfid}
 \end{center}
\end{figure}

A measurement of the integrated and differential fiducial production cross sections
for the Higgs boson in the diphoton decay channel at $\sqrt{s}=13~\mathrm{TeV}$
is performed using $35.9~\mathrm{fb}^{-1}$ of pp collisions data.
Events are split into categories of the mass resolution $\sigmaMoMdecorr$ to maximize the analysis sensitivity
to the SM Higgs boson. Differential cross sections are measured as a function of the diphoton transverse momentum
and jet multiplicity.
All cross sections are measured within a fiducial phase space defined by the requirements on the photons kinematics, their isolation, and the event topology. The measured cross sections are compared to state of the art theoretical predictions for the Standard Model Higgs bosons. A good agreement between observations and predictions is observed, as shown in Fig.~\ref{fig:Hggfid} .

\section{$\mathrm{H}\rightarrow ZZ $}
\label{sec:HZZ}

The $\HZZfl$ decay channel ($\ell=\Pe,\Pgm$) has a large signal-to-background ratio, and the precise
reconstruction of the final-state decay products allows the complete determination of the
kinematics of the Higgs boson. This makes it one of the most important channels for studies of the Higgs boson's properties.
The measurements of properties of the Higgs boson in the $\HZZfl$ decay channel at $\sqrt{s}=13\TeV$ are summarized.
Categories have been introduced targeting subleading production modes
of the Higgs boson such as vector boson fusion (VBF) and associated production with a vector boson ($\WH$, $\ZH$)
or top quark pair ($\ttH$).
In addition, dedicated measurements of the boson's mass, width, total and differential cross sections have been performed.

The full kinematic information from each event using either the Higgs boson decay products or
associated particles in its production is extracted using matrix element calculations and used
to form several kinematic discriminants.
The discriminant sensitive to the $\Pg\Pg/q\bar{q}\to4\ell$ kinematics, $\mathcal{D}^{\rm kin}_{\rm bkg}$,
is calculated as~\cite{Chatrchyan:2012xdj,Khachatryan:2014kca}. Four discriminants calculated as~\cite{Khachatryan:2015cwa,Khachatryan:2015mma}
are used to enhance the purity of event categories which are sensitive to the VBF signal topology with two associated jets ($\DMeVbfjj$),
the VBF signal topology with one associated jet (\DMeVbfj), and to the VH (either ZH or WH) signal topology with two associated jets ($\DMeWh$ and $\DMeZh$).

In order to improve the sensitivity to the Higgs boson production mechanisms,
the selected events are classified into seven categories,
based on the multiplicity of jets, b-tagged jets and additional leptons, missing energy and selections on kinematic discriminants $\mathcal{D}^{\rm kin}_{\rm bkg}$.
The seven categories are VBF-2jet-tagged category, VH-hadronic-tagged category, VH-leptonic-tagged category, ttH-tagged category, VH-MET-tagged category, VBF-1jet-tagged category and untagged category.

The irreducible background to the Higgs boson signal in the $4\ell$ channel, which come from the
production of $\cPZ\cPZ$ via $\Pq\Paq$ annihilation or gluon fusion, is estimated using simulation.
Additional backgrounds to the Higgs boson signal in the $4\ell$ channel arise
from processes in which heavy-flavor jets produce secondary leptons, and
also from processes in which decays of heavy-flavor hadrons, in-flight decays of
light mesons within jets, or (for electrons) the decay of charged
hadrons overlapping with $\pi^0$ decays are misidentified as leptons.
The main processes producing these backgrounds are $\cPZ+{\rm jets}$,  $\ttbar+{\rm jets}$,
$\cPZ\gamma+{\rm jets}$, $\PW\PW+{\rm jets}$, and $\PW\cPZ+{\rm jets}$.
We denote these reducible backgrounds as ``Z+X'' since they are dominated
by the $\cPZ+{\rm jets}$ process.

\begin{figure}
 \begin{center}
   \includegraphics[width=0.325\textwidth]{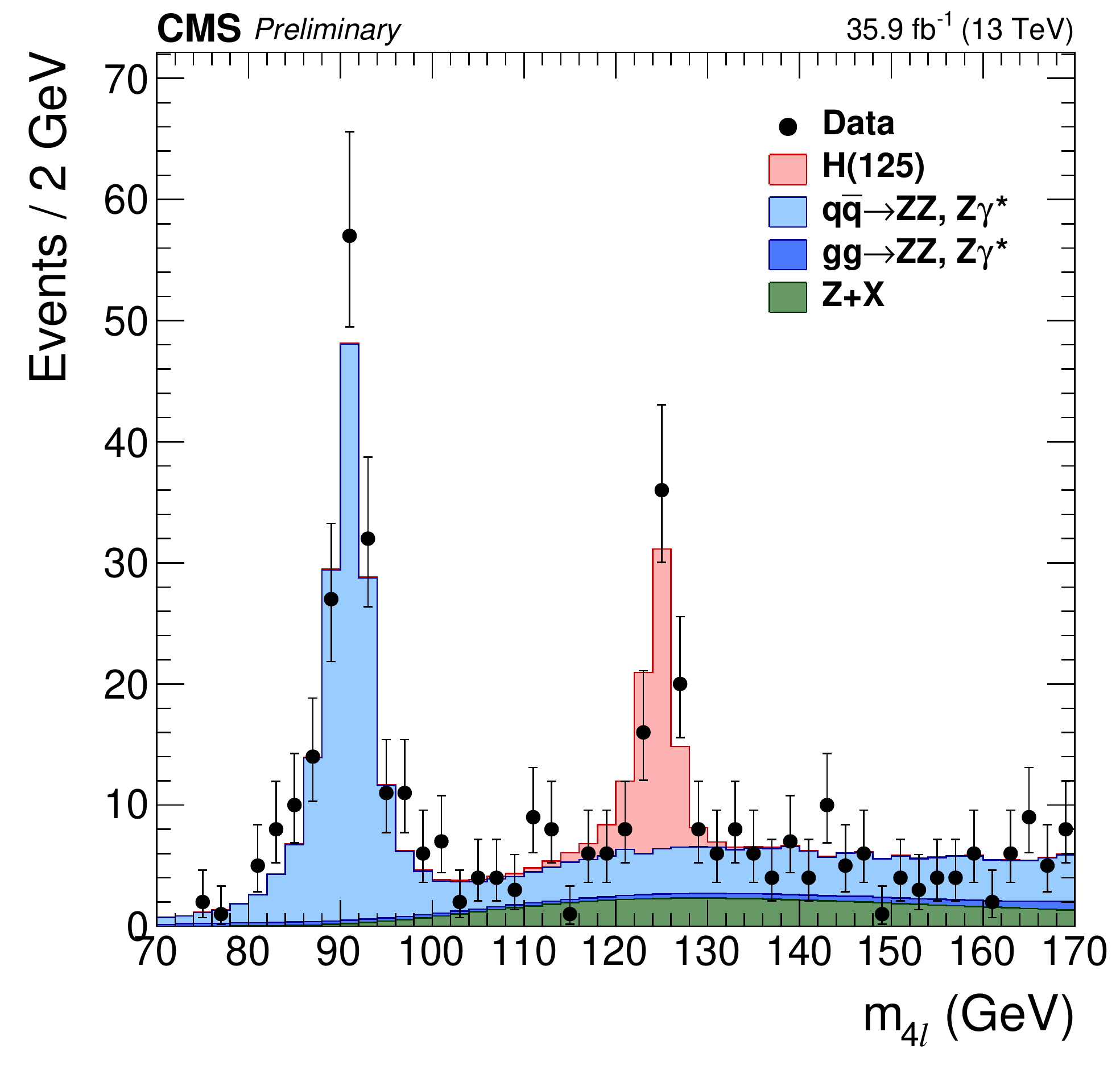}
   \includegraphics[width=0.325\textwidth]{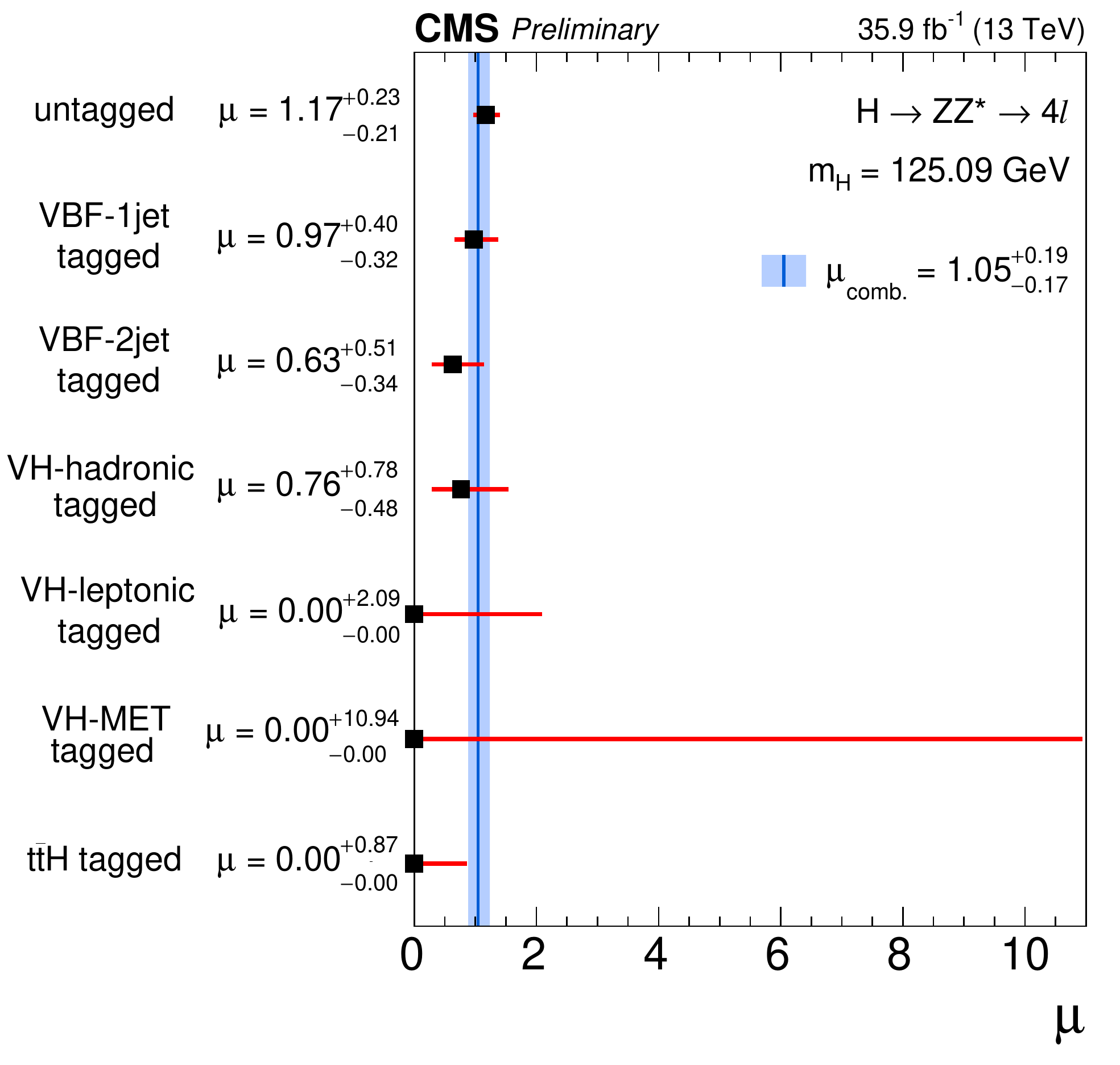}
   \includegraphics[width=0.325\textwidth]{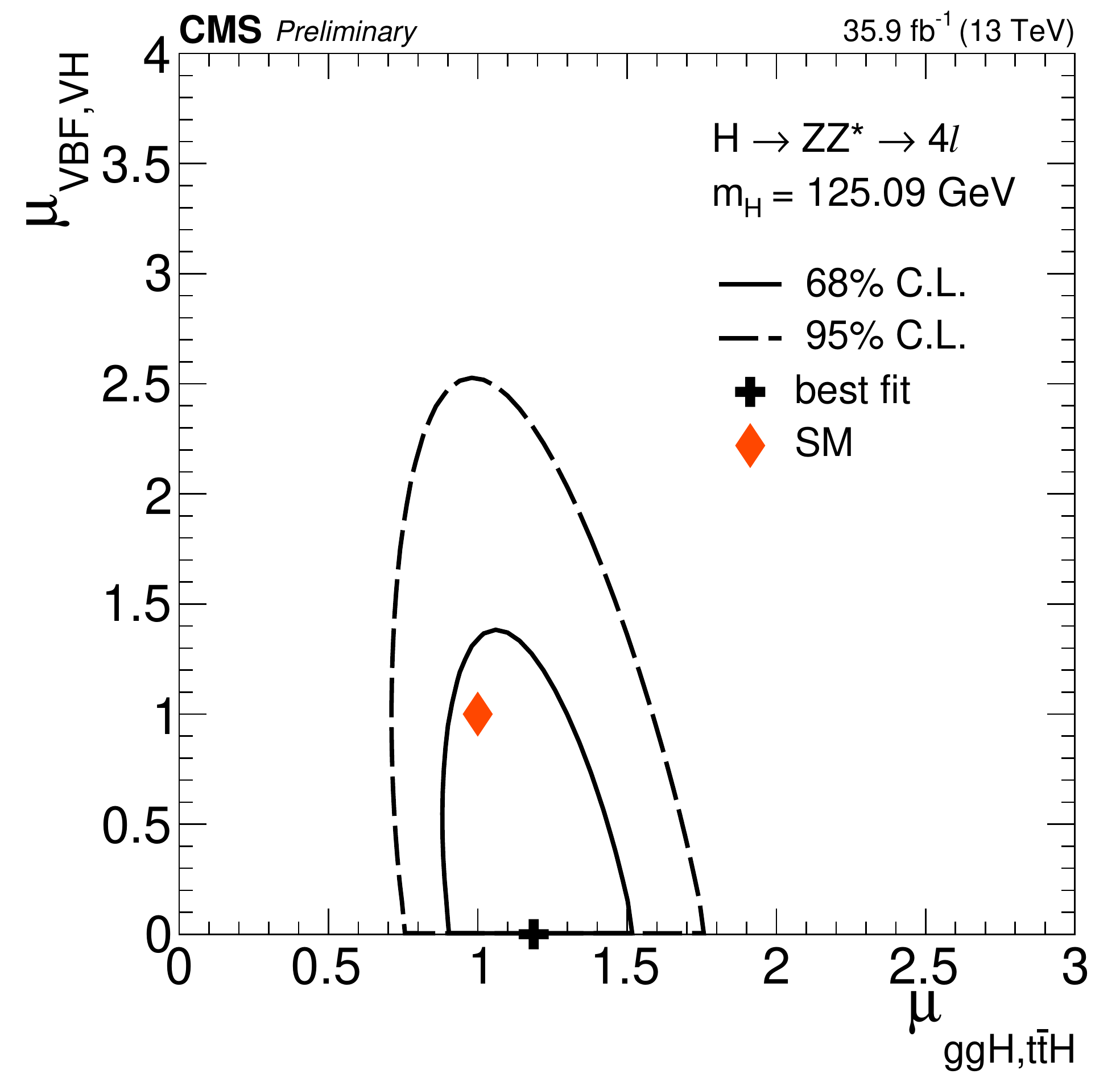}
   \caption{(Left) Distribution of the four-lepton reconstructed invariant mass $\mllll$ in the low-mass range. (Middle) Observed values of the signal strength $\mu=\sigma/\sigma_{SM}$ for the seven event categories, compared to the combined $\mu$ shown as a vertical line. The horizontal bars and the filled band indicate the $\pm$ 1$\sigma$ uncertainties. (Right) Result of the 2D likelihood scan for the $\muF$ and $\muV$ signal-strength modifiers.
The solid and dashed contours show the 68\% and 95\% CL regions, respectively.
The cross indicates the best-fit values, and the diamond represents the expected values for the SM Higgs boson.}
  \label{fig:HZZ4l}
 \end{center}
\end{figure}

The reconstructed four-lepton invariant mass distribution is shown in Fig.~\ref{fig:HZZ4l} for the sum of the  $4\Pe$, $4\Pgm$ and $2\Pe2\Pgm$ subchannels, and compared with the expectations from signal and background processes. A simultaneous fit to all categories is performed to extract the signal-strength modifier,
defined as the measured cross section relative to the expectation for the SM Higgs boson.
The fit relies on two variables: the four-lepton invariant mass $\mllll$ and the $\KD$ discriminant.
At $\mH=125.09\GeV$, the combined result is $\mu = \sigma/\sigma_{SM} = 1.05^{+0.15}_{-0.14}({\rm stat.})^{+0.11}_{-0.09}({\rm sys.}) =\valMuAtRunIMass$, which is compared to the results for each of the seven event categories in Fig.~\ref{fig:HZZ4l} (middle).
The observed value is consistent with 1 within the uncertainties.
Two signal-strength modifiers $\muF$ and $\muV$ are introduced as scale factors for the fermion and vector-boson induced contribution to the expected SM cross section.
A two-dimensional fit is performed assuming a mass of $\mH = 125.09\GeV$ leading to the measurements of  $\muF=\valMuFAtRunIMass$ and $\muV=\valMuVAtRunIMass$.
The 68\% and 95\% CL contours in the ($\muF,\muV$) plane are shown in Fig.~\ref{fig:HZZ4l} (right).

\begin{figure}
 \begin{center}
   \includegraphics[width=0.325\textwidth]{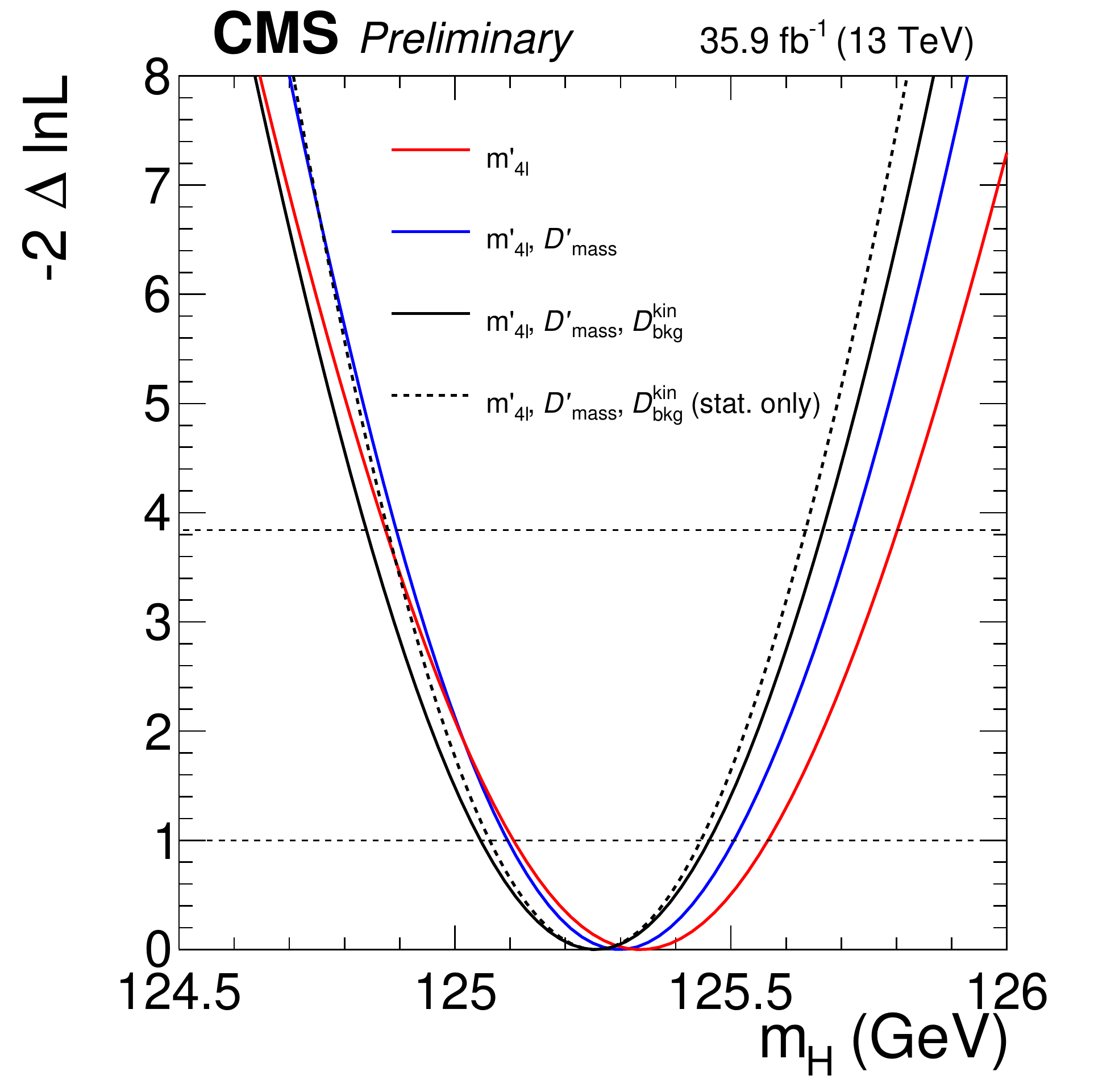}
   \includegraphics[width=0.325\textwidth]{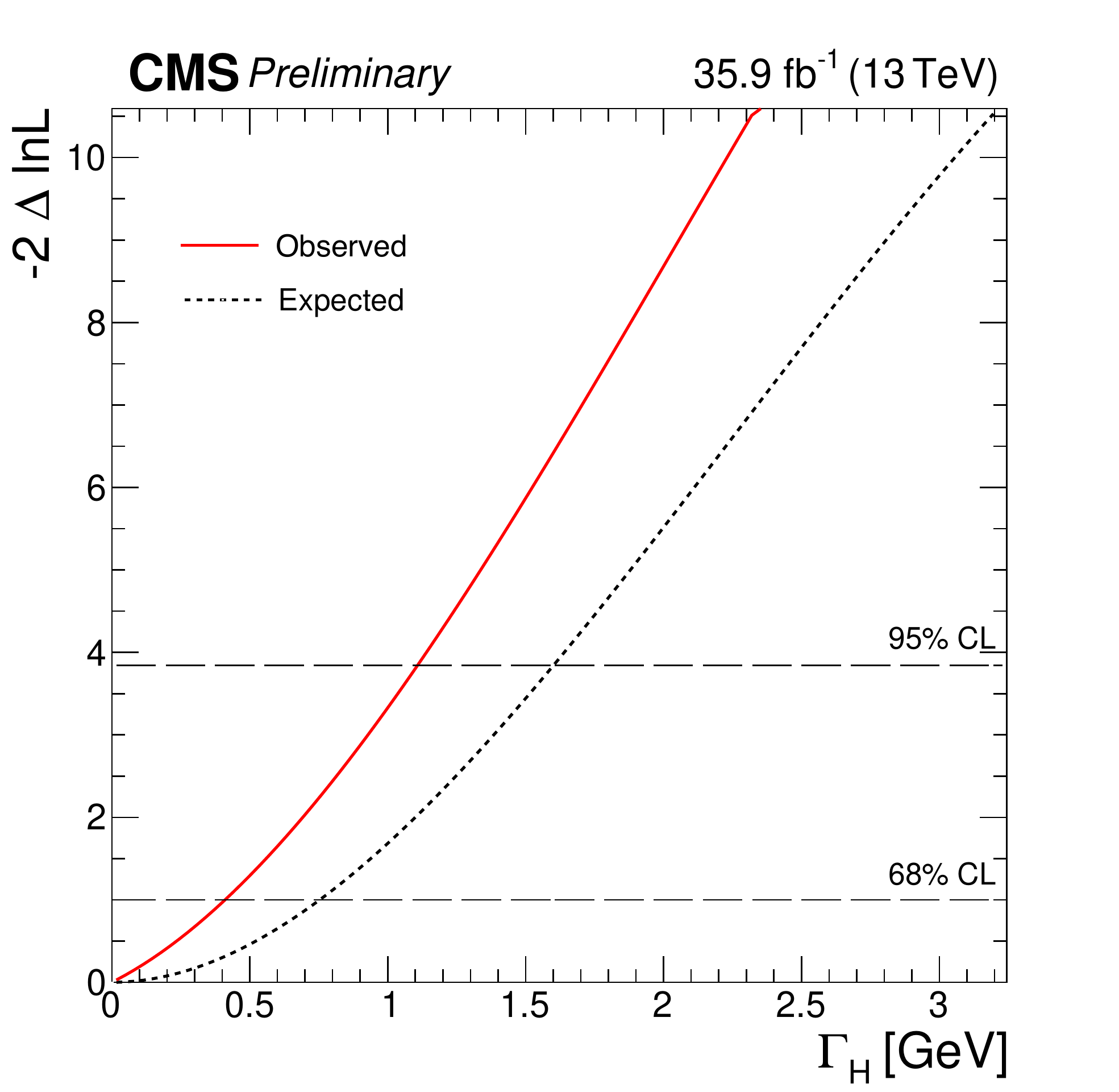}
   \includegraphics[width=0.325\textwidth]{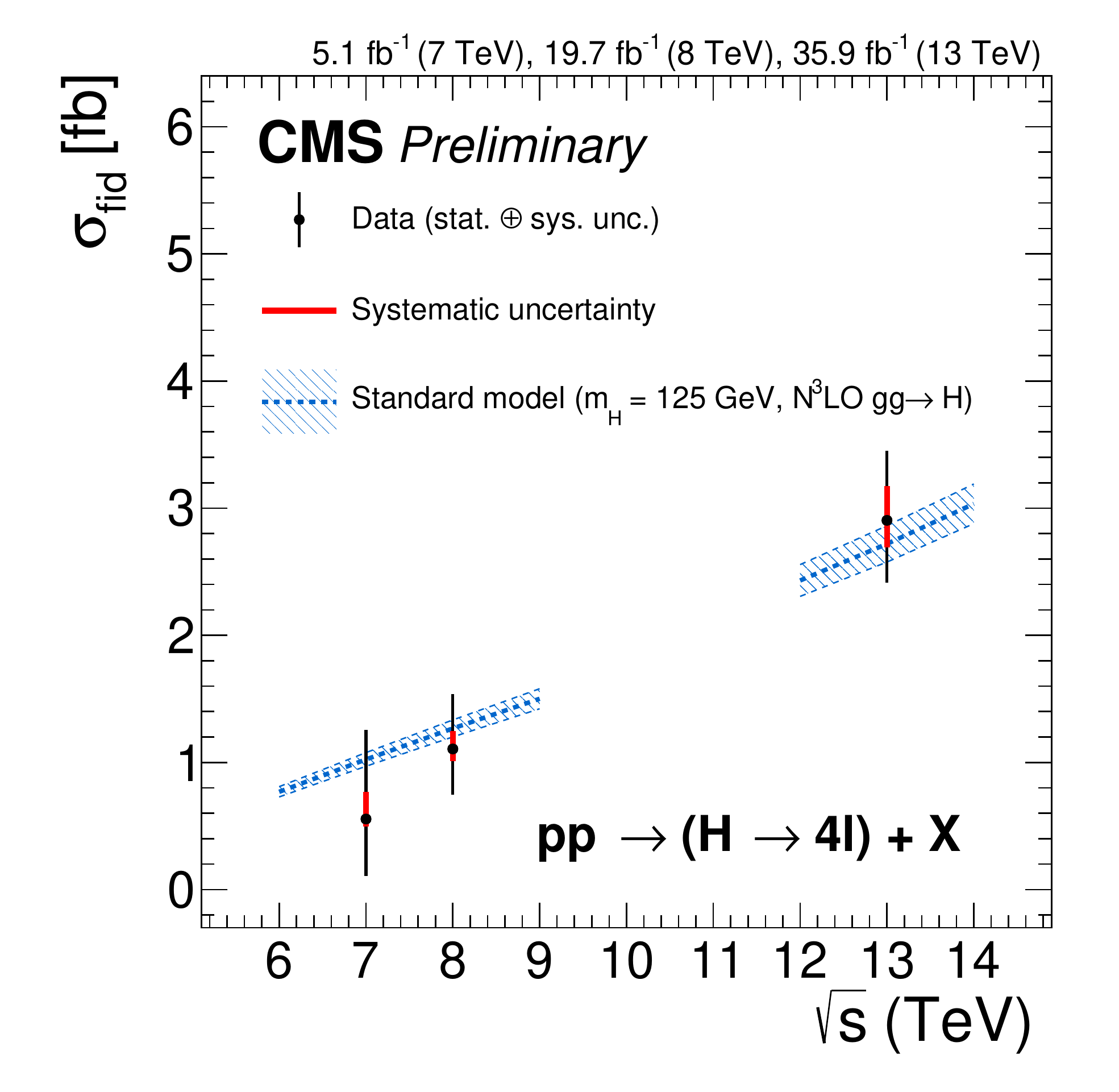}
   \caption{(Left) 1D likelihood scan as a function of mass for the 1D, 2D, and 3D measurement. Solid lines represent the scan with full
      uncertainties included, dashed lines statistical uncertainty only. (Middle) Observed and expected likelihood scan of $\Gamma_{\rm H}$ using the signal range $105 < m_{4\ell} < 140$~GeV, with $m_{\rm H}$ floated. (Right) The measured fiducial cross section as a function of $\sqrt{s}$. }
  \label{fig:HZZ4l2}
 \end{center}
\end{figure}

\begin{figure}
 \begin{center}
   \includegraphics[width=0.325\textwidth]{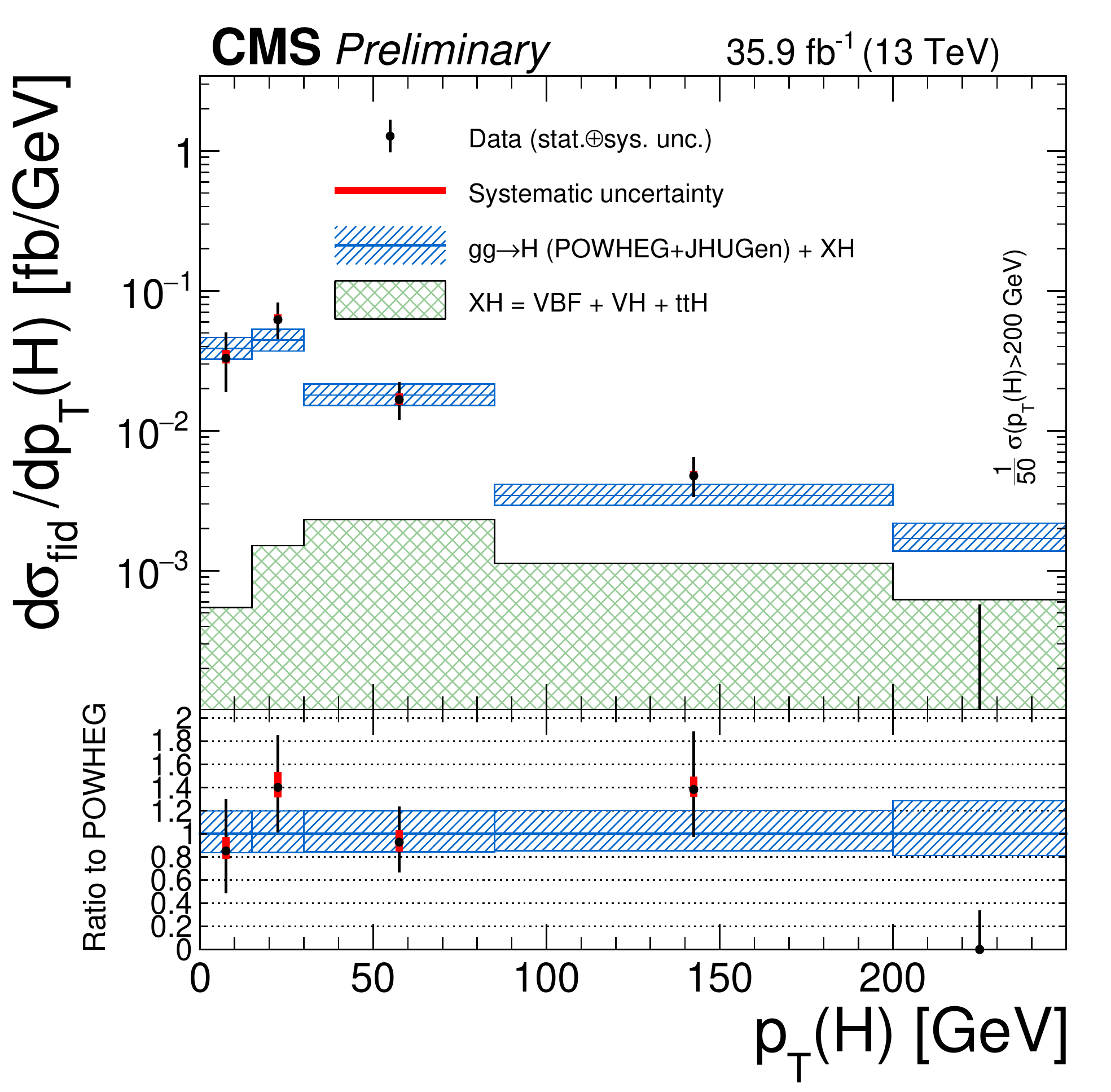}
   \includegraphics[width=0.325\textwidth]{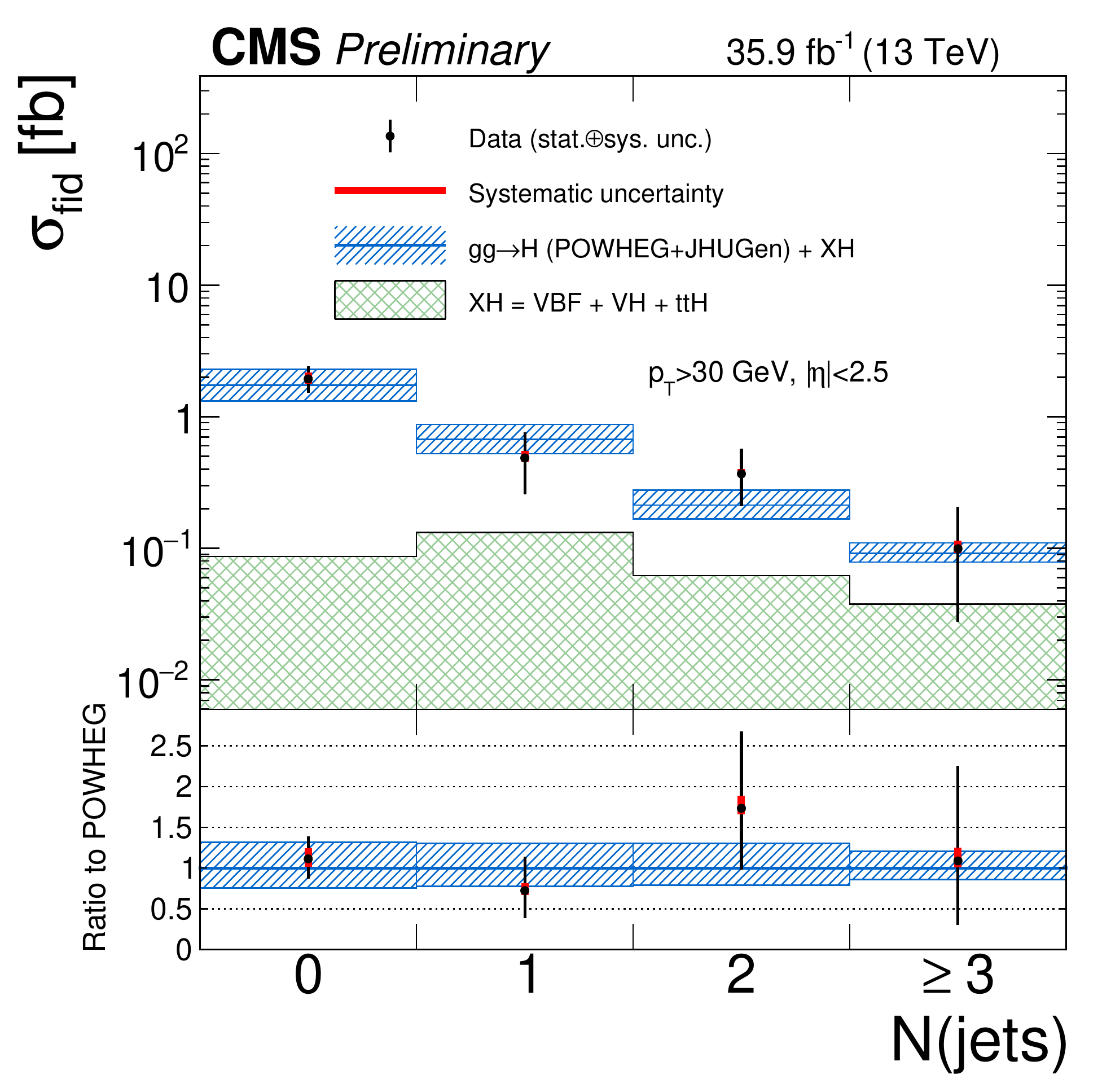}
   \includegraphics[width=0.325\textwidth]{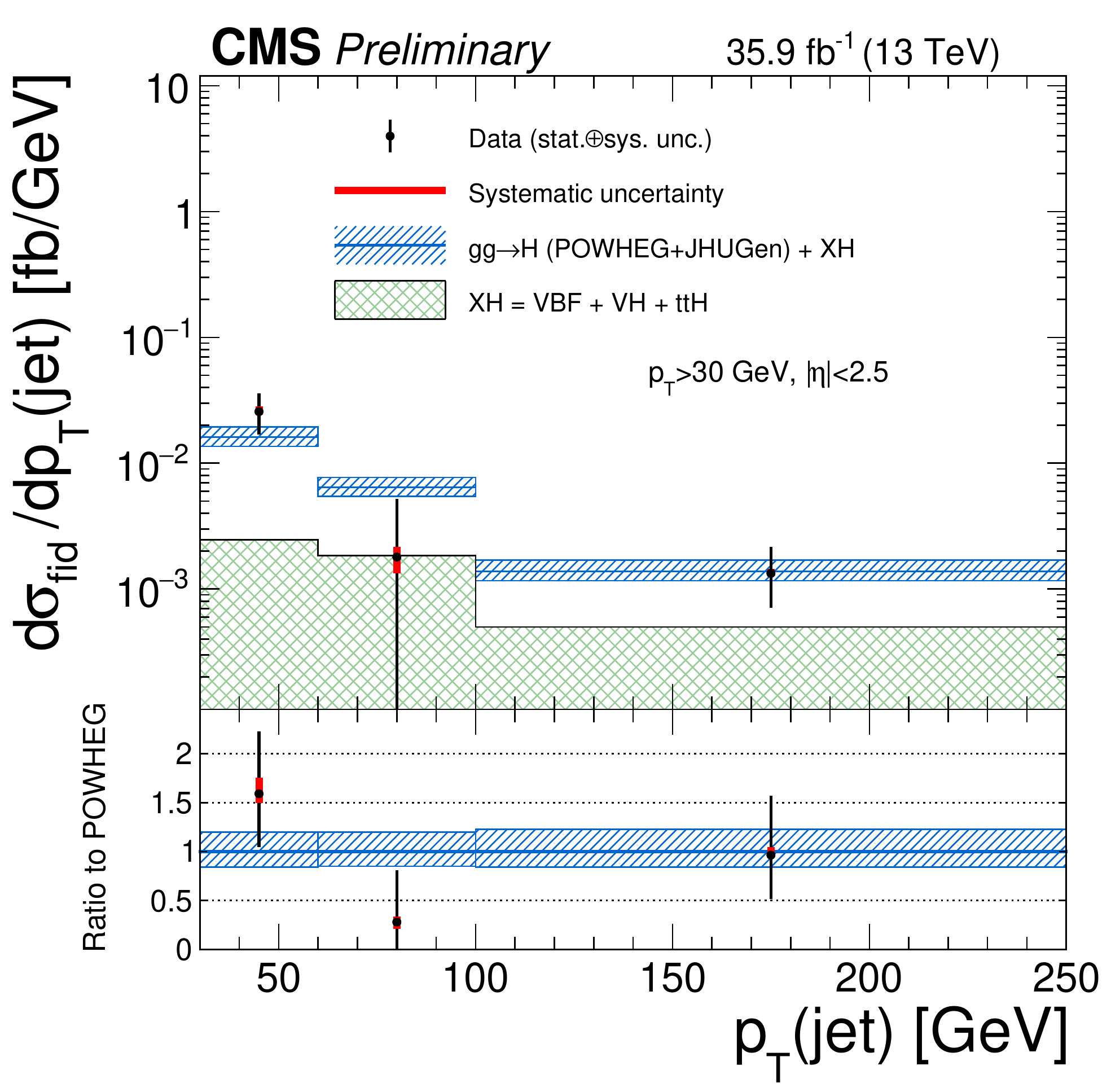}
   \caption{The results of the differential cross section measurement for $\pt({\rm H})$ (left), N(jets) (middle) and
$\pt({\rm jet})$ (right). The acceptance and theoretical uncertainties in the differential bins are calculated using POWHEG.
The sub-dominant component of the signal (VBF $+$ VH $+~\ttH$) is denoted as XH.}
  \label{fig:HZZDiferentialXS}
 \end{center}
\end{figure}

The 1-dimensional likelihood scans vs. $\mH$, while profiling the signal strength
$\mu$ along with all other nuisance parameters for the 1D ${\cal L}(m_{4l}')$ (four-lepton mass $m_{4l}'$), 2D ${\cal L}(m_{4l}',\MassDprime)$
 (four-lepton mass uncertainty $\MassDprime$) and 3D ${\cal L}(m_{4l}',\MassDprime,\KD)$ fits including the $m(\cPZ_1)$ constraint
are shown in Fig.~\ref{fig:HZZ4l2} (left). The best fitted value of $m_{\rm H}$ from 3D fit is $\valMassThreeDRefit~\GeV$.
Higgs width was measured using on-shell Higgs boson production,
using the range $105 < m_{4\ell} < 140$~GeV.
Figure~\ref{fig:HZZ4l2} (middle) shows the likelihood as a function of $\Gamma_{\rm H}$ with the $m_{\rm H}$ parameter unconstrained.
At 95\%~CL, the width $\Gamma_{\rm H}$  of the Higgs boson is less than 1.10 $\GeV$ (observed) and 1.60 $\GeV$ (expected).
The integrated fiducial cross section for the production and decay $\Pp\Pp\to\Hllll$ within a
fiducial volume defined to match closely the reconstruction level selection is measured to be
$\sigma_{{\rm fid.}}=2.90^{+0.48}_{-0.44}({\rm stat.})^{+0.27}_{-0.22}({\rm sys.})~{\rm fb}$.
This can be compared to the SM expectation $\sigma_{{\rm fid.}}^{\rm SM}=2.72\pm0.14~{\rm fb}$.
The integrated fiducial cross section as a function of $\sqrt{s}$ is also shown in Fig.~\ref{fig:HZZ4l2} (right).
The measured differential cross
section results for $\pt({\rm H})$, N(jets), and $\pt({\rm jet})$ can also be seen in Fig.~\ref{fig:HZZDiferentialXS}.

The study of the anomalous interactions of the Higgs boson is also performed using the decay information $\mathrm{H}\to 4\ell$ and information from associated production of two quark jets, originating either from vector boson fusion or associated vector boson. The tensor structure of the interactions of the spin-zero Higgs boson with two vector bosons either in production or in decay is investigated and constraints are set on anomalous HVV interactions. All observations are consistent with the expectations for the standard model Higgs boson.

\section{Summary}

The measurements of several properties of the standard model Higgs boson in both
the $\mathrm{H}\rightarrow\gamma\gamma$ decay channel and
the $\mathrm{H}\rightarrow{\rm Z}{\rm Z}\rightarrow4\ell$ ($\ell={\rm e},\mu$)
decay channel using data samples corresponding to an integrated luminosity of $\usedLumi$ at $\sqrt{s}=13~\TeV$ have been presented. The mass is measured to be $m_{{\rm H}}=\valMassThreeDRefit~$GeV  from the four-lepton final state.
All results are consistent, within their uncertainties, with the expectations for the SM Higgs boson.

\Acknowledgements

I would like to thank the LHCP2017 organizers for their hospitality and
the wonderful working environment. I acknowledge the support from
National Natural Science Foundation of China (No. 11661141007, No. 11061140514 and No. 11505208),
China Ministry of Science and Technology (No. 2013CB838700).

\end{document}